# About evaluation of radius of compactification and vacuum energy in the Randall-Sundrum model.

Sergey Yakovlev

Moscow, May 20, 2010
y_sergey@yahoo.com

Abstract

Using the base of the RS-model by introducing effective terms in common five-dimensional action an equation for radius of fifth dimension through cosmological constant and Planck mass was defined. Such a model have not a RS limit cause of the dependence vacuum energy from radius of additional dimension.

A standard approach for evaluation of radius of additional dimension is built on considering gravitational action for higher dimensional theories, that is given by following manner:

$$S = \frac{1}{16\pi G_{(D)}} \int d^D X \sqrt{-G^{(D)}} R^{(D)} ,$$

where $G_{(D)} = \frac{1}{M^{D-2}} \equiv \frac{1}{M^{d+2}}$, $d = D - 4$;

$$G^{(D)} = \det\|G_{MN}(X)\|, \quad M, N = 0,..., D-1,$$

$G_{(D)}$ – fundamental $D$-dimension Newton's constant, $d$ - is the number of extra dimensions, $G_{MN}(X)$ - metric tensor of $D$-dimensional space-time, $R^{(D)}(X)$ - scalar curvative, $X = \{X^M\}$ - coordinates in $D$ - dimensional space-time.

Four-dimensional effective gravitational action is getting after integration over additional dimensions, it is possible for the interesting matters with the diagonal metric, and we have

$$S_{eff} = \frac{V_d}{16\pi G_{(D)}} \int d^4 x \sqrt{-g^{(4)}} R^{(4)} ,$$

where $V_d \approx r^d$ is volume of additional dimensions. We see, that four dimensional Planck mass, with accuracy up to about numerical value about one, is equal

$$M_{PL} = M(Mr)^{d/2} .$$

If the size of additional dimension is more than fundamental size $M^{-1}$, Planck mass is more than fundamental gravitational scale $M$. Problem of mass hierarchy, as it turned out, connected with the size of extra dimensions.

For example, if we take $M = 1TeV$, we could calculate next evaluation for radius $r$ [3]:

$$r \approx M^{-1}\left(\frac{M_{PL}}{M}\right)^{2/d} \approx 10^{32/d} 10^{-19} m.$$

For one extra dimension we have that $r \approx 10^{13} m$. Evaluation for $M = 30TeV$ would give us that $r$ should be around size of $10^{-6} m$.

Let's try to get an expression for $r$ using the exactly solving Randall-Sundrum model. Write down standard action for the model, consisting of three components:

$$S = S_{GR} + S^+ + S^-,$$

$$S_{GR} = \int d^4x \int d\phi \sqrt{-G^{(5)}}(-\Lambda + 2M^3 R),$$

$$S^+ = -\int d^4x \sqrt{-g^+}\, V^+, \quad S^- = -\int d^4x \sqrt{-g^-}\, V^-,$$

$$g^{+-} = \det\|g_{\alpha\beta}^{+-}(x)\|; \quad \phi \in [-\pi,\pi]; \quad \alpha,\beta = 0,1,2,3; \tag{1}$$

here sign «+» corresponds to the brane with $\phi = 0$ and «-» - to the brane with $\phi = \pi$; first term in (1) is typical gravitation action with cosmological constant $\Lambda$, but second and third ones is actions of the two branes, where we took only vacuum energies $V^{+-}$ and where $g_{\alpha\beta}^{+-}(x)$ - induced metric on the branes for corresponding coordinate $\phi$.

Let's introduce action with following additional terms $I^+$ and $I^-$, that compensate tension on the branes $\Lambda$ while we integrate over branes is second expression in (1):

$$I^+ = -\int d^5 X \sqrt{-G^{(5)}}(-\Lambda)\delta(\phi), \quad I^- = -\int d^5 X \sqrt{-G^{(5)}}(-\Lambda)\delta(\phi - \pi),$$

or,

$$I^{+-} = -r\int d^4x \sqrt{-g^{+-}}(-\Lambda),$$

cause metric in (1) is taken in classical form [1]

$$ds^2 = e^{2\sigma(\phi)}\eta_{\mu\nu}dx^\mu dx^\nu + r^2 d\phi^2.$$

Then effective action will get the next form, where density of vacuum energy is changed to effective terms in little round brackets

$$S_{eff} = S_{GR} - \int d^4x \left[ \sqrt{-g^+}(V^+ - r\Lambda) + \sqrt{-g^-}(V^- - r\Lambda) \right]. \qquad (2)$$

Einstein's equation we could write down in form:

$$\sqrt{-G^{(5)}}(R_{MN} - \frac{1}{2}G_{MN}R) = -\frac{1}{4M^3}(\Lambda\sqrt{-G^{(5)}}G_{MN} + \sqrt{-g^+}(V^+ - r\Lambda)g^+_{\mu\nu}\delta^\mu_M\delta^\nu_N\delta(\phi) +$$

$$\sqrt{-g^-}(V^- - r\Lambda)g^-_{\mu\nu}\delta^\mu_M\delta^\nu_N\delta(\phi - \pi). \qquad (3)$$

The equations following from Eq. (3) by ordinary way [1] reduce to

$$\frac{6\sigma'^2}{r^2} = \frac{-\Lambda}{4M^3},$$

$$\frac{3\sigma''}{r^2} = \frac{(V^+ - r\Lambda)}{4M^3 r}\delta(\phi) + \frac{(V^- - r\Lambda)}{4M^3 r}\delta(\phi - \pi).$$

From here we obtain following equations for the metric (2):

$$V^- - r\Lambda = -\sqrt{-24M^3\Lambda},$$

$$V^+ - r\Lambda = \sqrt{-24M^3\Lambda},$$

$$\sigma(\phi) = -r\sqrt{-\frac{\Lambda}{24M^3}}|\phi|, \qquad \Lambda < 0.$$

Observing $r$ and $\Lambda$ through $V^+, V^-$ we have

$$r = -\frac{48(V^+ + V^-)M^3}{(V^+ - V^-)^2}, \quad V^+ + V^- \leq 0;$$

$$\Lambda = -\frac{(V^+ - V^-)^2}{96M^3},$$

$$\sigma(\phi) = \frac{(V^+ + V^-)}{(V^+ - V^-)}|\phi|.$$

(4)

From this we see, that if $V^+ = -V^-$, five-dimensional space-time is reducing to four-dimensional, $r = 0$. We'll not consider here physical sense and consequences of such a result, considering only geometric content.

In special case, while we take tension on the brane $V^+ = 0$, from (3) we get

$$V^- = 2r\Lambda,$$

and value of radius of compactification $r$ is equal

$$r_0 = \left(-\frac{24M^3}{\Lambda}\right)^{1/2}.$$

And we have the metric in the next form:

$$ds^2 = e^{2\sigma(\phi)}\eta_{\mu\nu}dx^\mu dx^\nu - \frac{24M^3}{\Lambda}d\phi^2,$$
$$\sigma(\phi) = -|\phi|.$$

(5)

Thus, the metric on the "negative" brane with vacuum energy $V^- - r_0\Lambda = r_0\Lambda$, when $\phi = \pi$, is flat:

$$ds^2 = e^{-2\pi}\eta_{\mu\nu}dx^\mu dx^\nu;$$

On the "positive" brane, where we 've got $\phi = 0$, accordingly we have got vacuum energy $V^+ - r_0\Lambda = -r_0\Lambda$, and also the same Minkowski metric,

$$ds^2 = \eta_{\mu\nu}dx^\mu dx^\nu.$$

We see that metric's component

$$G_{44} = r_0^2 = -\frac{24M^3}{\Lambda},$$

as it's turned out, is inversely of value of cosmological constant of five-dimensional space-time, and signature of metric as turned out is standard; the signs of space components of metric are positive $(-++++)$ because of the cosmological constant, as we have it is negative. Of course, the real length of additional dimension is equal $\pi r_0$. Would be interested to consider metrics like (5) in general case.

Thus, in special case for the model where we took one of the branes $V^+ = 0$, we derived a clear expression for the radius of fifth dimension and metric (5). It's value is inverse of the square root of cosmological constant of fifth-dimensional space-time. We see, that in such an approach vacuum energy, or the same, cosmological constant could take the negative value, as in the classical Randall-Sundrum theory, that provides us the right signature of global metric and existence of solution of Einstein's equation.

The author is grateful to Neil Lessman for helpful communications regarding the translation of the manuscript.

**References:**

1. A large Mass Hierarchy from a Small Extra Dimension. Lisa Randall, Raman Sundrum, Phys.Rev.Lett. 83, 3370(1999) [hep-th/9905221].
2. An Alternative to Compactification. Lisa Randall, Raman Sundrum, Phys.Rev.Lett. 83, 4690(1999) [hep-th/9906064].
3. Large and infinite extra dimensions. V.A.Rubakov, Phys.Usp.44:871-893, 2001 [hep-ph/0104152].
4. U(1) gauge field of the Kaluza-Klein theory in the presence of branes. Gundwon Kang, Y.S. Myung, Phys.Rev.D63:064036, 2001 [hep-th/0007197].
5. Little Randall-Sundrum model and a multiply warped spacetime. Kristian L.McDonald. [hep-th/0804.0654v2], 10 Apr.2008.
6. Mathematical theory of black holes. S.Chandrasechar. Moscow, 1986.
7. Introduction to the theory of quantum fields. N.N.Bogolubov, D.V.Shirkov. Moscow, 1976.
8. Gravitation. C.W. Misner, K.S. Thorne, J.A. Wheeler, Moscow, 1996.
9. Mathematical theory of black holes. S.Chandrasechar. Moscow, 1986
10. Structure of space-time. R.Penrose. New York-Amsterdam, 1968.
11. Conserved currents in D-dimensional gravity and brane cosmology. A.N.Petrov. Arxive: gr-qc/0401085v2.